# Taxonomy of Virtual and Augmented Reality Applications in Education

Jiri Motejlek and Esat Alpay

**Abstract**

This paper presents and analyses existing taxonomies of virtual and augmented reality and demonstrates knowledge gaps and mixed terminology which may cause confusion among educators, researchers, and developers. Several such occasions of confusion are presented. A methodology is then presented to construct a taxonomy of virtual reality and augmented reality applications based on a combination of: a faceted analysis approach for the overall design of the taxonomy; an existing taxonomy of educational objectives to derive the educational purpose; an information systems analysis to establish important facets of the taxonomy; and two systematic mapping studies to identify categories within each facet. Using this methodology a new taxonomy is proposed and the implications of its facets (and combinations of facets) demonstrated. The taxonomy focuses on technology used to provide the virtual or augmented reality as well as the content presented to the user, including the type of gamification and how it is operated. It also accommodates a large number of devices and approaches developed throughout the years and for multiple industries, and provides a way to categorize them in order to clarify communication between researchers, developers and educators. Use of the taxonomy is then demonstrated in two case studies: a virtual reality chemical plant for use in chemical engineering education and an augmented reality dog for veterinary education.

## 1 Introduction

The importance of education in the social and economic development of society is one of the main drivers behind new initiatives to improve its impact. In recent years an introduction of relatively cheap hardware for virtual reality (VR) and augmented reality (AR) Brown and Green 2016, as well advancements in creative software to develop applications for these devices (Smith, Walford, and Jimenez-Bescos, 2019), has caused an increase in interest in VR and AR in educational contexts (Douglas-Lenders, Holland, and Allen, 2017). Both technologies have in most cases shown success (Shen et al., Apr. 3–7, 2017). Examples include military training applications (Pallavicini et al., 2013), engineering applications through VR laboratories (Ren et al., 2015) and education in the fields of history (Barreau et al., 2015) and astronomy (Yen, Tsai, and Wu, Nov. 26, 2013). The possibilities to use VR/AR transcend to other contexts, such as interactive performances, theatre, galleries, discovery centers, and so on (Schlacht, Mastro, and Nazir, Jul. 27-31, 2016). Greater understanding is needed as to the features of such applications that are especially conducive to student learning. Clarity is also needed on the classification of VR and AR applications to allow accurate description and comparison of research studies and, more generally, communication between researchers and developers. In this work, after an evaluation of three existing taxonomies (Vergara's, Hugues' and Milgram's (Hugues, Fuchs, and Nannipieri, 2011; Milgram and Kishino, 1994; Vergara, Rubio, and Lorenzo, 2017)) knowledge gaps and potential sources of confusion for educators, researchers, and developers, are described. Subsequently, a methodology aiming to construct a new taxonomy of VR/AR applications is presented. The methodology described uses a combination of a faceted analysis approach (Kwasnik, 1999), an existing taxonomy of educational objectives (Bloom et al., 1956) and information systems analysis (Eeles, 2001). The methodology is then used to construct a new taxonomy of VR/AR applications and the implications of some of the taxonomy facets (and their combinations) are demonstrated through two case studies. The work extends on a basic introduction to the requirements of the a taxonomy that was presented at the 2018 annual conference of the European Society for Engineering Education (Motejlek and Alpay, 2018).

## 2 Background

Three existing taxonomies of VR and AR technologies were evaluated: Vergara's taxonomy (Vergara, Rubio, and Lorenzo, 2017), Hugues' taxonomy (Hugues, Fuchs, and Nannipieri, 2011), and Milgram's taxonomy (Milgram and Kishino, 1994). According to the taxonomy developed by Vergara (Vergara, Rubio, and Lorenzo, 2017) for VR, educational VR applications can be categorized by the level of immersion as: 1) non-immersive (where the user's vision is as a "window in the world" displayed on a computer screen), and 2) immersive (which completely introduces the user into the virtual world, using for example a head-mounted display [HMD]). Vergara's work further specifies that immersive VR applications divide into two subcategories, HMD and room-scale CAVE (i.e., a setupinvolving projectors displaying synchronised images on the walls of a room ), and non-immersive VR applications divide into those that are operated with conventional peripherals and those using special devices. This taxonomy is not complete in its own domain, lacking a computer screen (e.g., in the case of driving simulators) or not taking into account the content displayed and its impact on immersion). Furthermore, no evidence is provided as to why immersion is considered a useful factor, as immersion is a qualitative result of many factors (such as interaction with the system or the type of content displayed). A number of papers describe immersive virtual worlds being displayed on computer screens (conventional peripherals), therefore the taxonomy does not help to categorize all existing research (see e.g., (Warburton, 2009)).

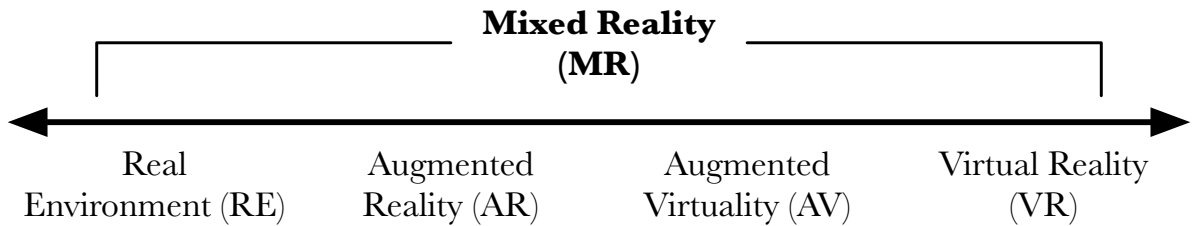

Figure 1: Virtuality continuum according to Milgram's taxonomy (Milgram and Kishino, 1994)

A taxonomy developed by Hugues (Hugues, Fuchs, and Nannipieri, 2011) focuses on AR. The taxonomy proposes that AR is a tool for assisting decision making by providing information for the better understanding reality. It does not rely on technology used to facilitate AR, but rather on the type of information shown to the user. The taxonomy divides AR into two major categories: artificial environments and augmented perception. The artificial environments category encompasses applications helping the user to envision what a reality could potentially look like, further differentiating between future, past and fictional reality. Augmented perception focuses on the method that AR can provide information for assisting decision making. Categories seem to overlap, for example both "augmented comprehension" (displaying virtual annotations on real objects) and "augmented visibility" (displaying virtual contours on real objects) present computer-generated data contextual to reality. The subcategories of "artificial environments" are contextually different to those in "augmented perception" without an apparent reason as to why a historical perspective was chosen in the former and a technological solution in the latter.

The taxonomy developed by Milgram (Milgram and Kishino, 1994) is developed in the form of a series of scales (continua) involving both AR and VR and focuses on the type of imaging device used to deliver the experience. The virtuality continuum according to the taxonomy is depicted in 1.

The idea behind Milgram's taxonomy is that one technology (AR) can continuously transform into another (VR). This has been accepted by many researchers, who then confuse the context of the word "virtual" as meaning both VR and AR. For example, using this taxonomy, Tepper (Tepper et al., 2017) concludes that the Microsoft HoloLens can display both VR and AR, which is not the case as the technologically cannot display the former. In order for Milgram's taxonomy to be useful, the way that VR and AR design (or application) is approached by educators and developers would also have to be on a similar continuous scale. It will be shown in 7.2 that these two experiences are in fact discreet in nature. Similarly, the reproduction fidelity used within the taxonomy is not useful as it highly depends on what is reproduced. For example, a simulation of microsurgery using a physical laparoscope and computer monitor has a high reproduction fidelity (almost unrecognizable from a real life laparoscope) but is achieved with a simple conventional monoscopic video (which according to the taxonomy should provide the worst results). Furthermore, the "extent of presence" in this taxonnomy connects the display device to imaging type, which do not necessarily depend on each other. For example, an HMD can be used to provide monoscopic imaging, and surrogate travel is possible in the case of large screens (e.g., CAVE (Bruder et al., 2016)), but both of these statements should be incorrect according to the taxonomy.

Other taxonomies exist, such as detailed classification of AR visualization methods and techniques Zollmann et al. 2020 and a taxonomy focusing on virtual worlds and their use in education Duncan, Miller, and Jiang 2012. These taxonomies are rather detailed but provide a precise language in their specific domains. As will be shown, confusion of terminology and meaning exists in other works on a more fundamental level.

Table 1: Research Questions.

| Research question | Main motivation |
|---|---|
| What are the key perspectives on VR/AR technology and software that are of particular use for educators, developers, and researchers? | To identify facets used to construct the taxonomy |
| What educational purpose can VR/AR applications have? | To identify and guide the decision making process of educators when designing or choosing VR/AR applications for use for educational purposes. |
| What VR/AR hardware is being currently used for education purposes? | To discover non-overlapping categories to group the hardware for each identified facet. |
| What VR/AR applications are being currently used for education purposes? | To discover non-overlapping categories to group the applications for each identified facet. |
| What are the practical implications of the taxonomy? | To demonstrate the usefulness of the taxonomy. |

## 2.1 Confusion and Errors in Terminology

The need for the new taxonomy stems from an apparent confusion amongst educators and researchers regarding VR/AR terminology. These are demonstrated by some of the examples summarized below:

1. Computer facilitated 3D environments are in essence virtual (as in, not real) and are by some considered to be a virtual reality (e.g., (Mikropoulos and Natsis, 2011)). In a widely cited article "Measuring Presence in Virtual Environments: A Presence Questionnaire" (Witmer and Singer, 1998) authors describe a method of measuring a presence in 3D environments that are generated on a computer. The authors did not intend to imply that the environments are viewed through an immersive VR headset (HMD), yet, the presence questionnaire is used as a reference in research only relevant to the immersive VR which involves headsets (such as (Farmani and Teather, 2020)), even though the emotional impact on the user differs between the two technologies, as shown in a number of studies (see e.g., (Chirico et al., 2018)). This example demonstrates that as a result of the recent development of VR technology, the terminology once perfectly correct can be misinterpreted.

2. Medical students are using an imitation of an endoscope coupled with a screen for training purposes, for example MIST VR (M. S. Wilson et al., 1997), but there are also simulators offering 3D capabilities (Mostafa et al., 2017). Clear differentiation between these cases is necessary.

3. According to Milgram's taxonomy, VR and AR are subsets of Mixed Reality. Based on this, Microsoft called their technology "Windows Mixed Reality", but in fact produced an AR headset. This confusion in terms is then repeated by other researchers (e.g., (Mekni and Lemieux, 2014)). According to Microsoft, their HoloLens is delivering mixed reality, and so by some considered to be capable of displaying both VR and AR (Tepper et al., 2017). This statement is incorrect as Microsoft HoloLens is not capable of displaying VR content (it is in fact not capable of displaying darker images than the user's environment). This shows how older taxonomies can misguide the selection of VR/AR capable devices.

4. Google glass is considered to be an AR device even though it cannot augment reality (it can only display a screen in front of the user with information unrelated to the user's environment). Displaying information does not automatically 'augment' wearers reality, even if the information is useful to the wearer. If we would accept this, we would have to accept that, for example, wrist watches are an example of AR devices. A recognition of the information being registered in the 3D space around the user is necessary (Azuma, 1997).

Scope exists for a new taxonomy to allow better scrutiny behind design principles during the development process by allowing clear communication between educators, researchers, and developers. Such a taxonomy would also simplify the evaluation of VR/AR applications by educators through focusing on educational purpose and avoiding application development without a clear goal or clear evaluation strategy (e.g., (Zheng and Waller, 2017)). Thus, based on the above status of VR and AR classification and interpretation, the specific research questions summarized in 1 arise.

# 3 Methodology

The proposed taxonomy has been structured around the principles of the faceted analysis approach (Kwasnik, 1999), which allows the taxonomy to be flexible (new categories can be added if needed), offer multiple perspectives (its use can focus on a particular category and its descriptors' meaning) and is expressive (allows for a variety of VR/AR experiences to be captured. The aim of the taxonomy is to help educators, developers, and researchers in educational technologies to fully appreciate the facets of, for example, VR/AR hardware, software, educational purpose, and functionality. For that reason:

1. Construction of the list of taxonomy facets was based on the FURPS+ methodology (Eeles, 2001) used in software engineering. FURPS+ methodology is a generic and widely used method of gathering requirements of projects in software engineering. It was chosen for its simplicity and intuitiveness whilst retaining usefulness. This helps the resulting taxonomy to be understandable and still follow the principles of information system design.

2. Domains of Bloom's taxonomy of educational objectives (Bloom et al., 1956) were used to generate a classification of Educational Purpose. The domains were selected for their simplicity, yet effective coverage of the fundamental purpose of an educational activity or experience, and the cognitive, affective and psychomotor facets of education (see 5 below).

3. Two systematic mapping studies were performed, both following the guidelines proposed by (Kitchenham and Charters, 2007): 1) a hardware systematic mapping study and 2) a software systematic mapping study.

4. Categories of each facet were identified from the descriptions of the technologies and of educational content. The categories of each classification were developed in order to comply with the following rules:

    (a) *Parsimonious*
        Containing only reasonably non-overlapping categories and no unnecessary categories.
    (b) *Comprehensive*
        Covering all the aspects of the area the given part of the taxonomy aims to cover.
    (c) *Useful*
        Ability to help with decision making regarding the development process of the VR experience.

In the final section of the papr, the use of the taxonomy is demonstrated and the implications of each category outlined.

# 4 Facets Identification

The research question "What are the key perspectives on VR/AR technology and software that are of particular use for educators, developers, and researchers?" is considered through a facets identification methodology. Facets identification is based on information systems design principles and systems requirements gathering as those typically form the initial phase of software analysis. Four concerns defined in the FURPS+ methodology (Eeles, 2001) were used to compile the main structure of this taxonomy:

1. *Design Requirements*

    Design requirements are specified by the educator and form an integral part of information gathering. They are critical to successful use of the application in educational settings, successful development and meaningful analysis. To identify the design requirements the following classification was developed:

    - *Educational Purpose* based on domains of a taxonomy of educational goals designed by Bloom (Bloom et al., 1956).
    - *Type of Experience* based on the extent of the user's isolation.

2. *Physical Requirements*

    Physical requirements specify the technology that is used to provide the experience to the user.

    - *Delivery Technology* defines which technology is used to directly present the VR/AR content to the user.

- *Back-End Technology* defines what powers the delivery technology to provide the VR/AR experience.

3. *Interface Requirements*

   Interface requirements specify how the system interacts with the user and vice versa.

   - *User Interaction* defines how the user can interact with the application.
   - *System Interaction* defines how the system communicates with the user in order to achieve educational objectives.

4. *Implementation Requirements*

   Implementation requirements define how the system is built.

   - *Production Technology* focuses on the technical implementation of production.
   - *Type of Gamification* focuses on how the educational goals are achieved by the application.

Each of the eight facets consists of a set of defining categories (see 2 as an illustration of the subsequent taxonomy structure - category selection details are described below). The aim of this taxonomy is that selection of one category per facet allows for a comprehensive description of the VR/AR experience (as will be demonstrated in 8). User interaction, system interaction and gamification are optional facets. fFor example, a VR application may contain no gamification at all.

# 5 Derivation of Educational Purpose

The research question "What educational purpose can VR/AR applications have" is considered through a facets identification methodology. The chosen methodology for categorizing the educational purpose of the taxonomy is based on Bloom's taxonomy of educational goals (Bloom et al., 1956). The purpose of the educational application is the key premise of the VR/AR application in this taxonomy. Activities constructed and performed with and around the VR/AR application consist of actions which are motivated by a desire to achieve specific goals (Leontyev, 1978), in this case learning outcomes. The ability to understand the purpose of the application by the user is important to increase motivation to use such VR/AR experience for learning (Beke Hen, 2019).

The taxonomy devised by Bloom (Bloom et al., 1956) is divided into three domains: the cognitive domain, affective domain and psychomotor domain. Each domain corresponds to a category under the Education Purpose facet (see 2), namely:

1. *Training*

   Based on the psychomotor domain: a very specific activity aiming at teaching the student a particular skill; the goal is for the student to be able to apply the skill.

2. *Teaching*

   Based on the cognitive domain: this represents knowledge transfer, aiming to teach the student a principle; the goal is for the student is to be able to recollect the principle.

3. *Observing*

   Based on the affective domain: this aims to engage the student and motivate the student to learn more about a specific subject.

To derive the categories for the remaining facets, two systematic mapping studies were conducted. Further description of these is given below.

# 6 Systematic Mapping Study

Two systematic mapping studies were conducted in order to identify hardware and software used in VR/AR. Guidelines proposed by (Kitchenham and Charters, 2007) were followed to develop a protocol in the planning phase of the study as done by other researchers (e.g., (Vargas et al., May, 2014)), which includes the identification of the research question, a search strategy, selection strategy, selection procedure, data extraction strategy and data synthesis. The journals selected focus on STEM. However, due to the commonality of the technologies employed, the taxonomy is expected to be universal in its application.

Search keywords were identified in the following way (Brereton et al., 2007):

Table 2: Search and Selection Strategy.

| | |
|---|---|
| Databases searched | ACM Digital Library, IEEE Digital Library, MIT Press, The Royal Society, Springer Database, PubMed, De Gruyter, ERIC |
| Target items | Journal papers, Workshop papers, Conference papers |
| Search applied to | Title, Abstract, Keywords |
| Language | English |
| Publication period | Until March 2020 |
| Inclusion criteria | Papers that fulfill the search string, Journal and conference papers, Papers written in English, Papers published until March 2020 |
| Exclusion criteria | Papers focusing on developing new technology |

- Derive main keywords from the research question
- Identify alternative keywords, synonyms and related keywords
- Use the Boolean OR to incorporate alternative spelling, synonyms and related keywords
- Use the Boolean AND to link the main keywords

The search and selection strategy for both studies was the same and the details can be seen in 2. No lower limit of publication period was applied in order to create a taxonomy which is universal enough to allow categorization of all VR/AR applications and which is not limited in scope by current trends.

## 6.1 Hardware Systematic Mapping Study

### 6.1.1 Study planning

As stated in the introduction section of this paper, the aim of this systematic mapping study was to discover non-overlapping categories which would allow grouping of the VR/AR hardware. The relevant research question related to this mapping study is "What VR/AR hardware is being currently used for education purposes?". Search terms derived from the research question and the alternative terms are in 3. The terms "virtual learning environments" and "VLE" were excluded.

Table 3: Hardware—Included Terms.

| Main included terms | Alternative terms |
|---|---|
| Virtual Reality, Virtual Worlds, Mixed Reality, Augmented Reality, Immersive Virtual Environments, Head-Mounted Display | VR, XR, AR, IVE, HMD, Telepresence |

In numerous articles, virtual worlds and virtual reality are used interchangeably (Warburton, 2009; Schultze, 2010), the term virtual reality originally includes virtual worlds (Steuer, 1992) therefore the term "virtual worlds" was included. A search was performed and appropriate papers were selected according to the selection strategy. Technologies / devices that provide a complete experience which authors describe as VR or AR were compiled into a list. Devices that are not a vital part of another VR/AR device were excluded and the device they attach to was included in the list instead.

### 6.1.2 Study results

Nineteen distinct devices were identified, with the following observations:

- Eight involve VR, six AR and five use a display.
- Ten of of the devices employed an HMD and nine a screen.
- Nine devices were stand alone, eight were mobile devices and one used a cloud back-end.

Based on the search terms "virtual reality", "virtual environments", "virtual worlds" and "augmented reality" a number of devices were selected, and by further removing devices with an identical function a list of unique example devices was compiled. Where the device provides the same functionality as another device in the list, the second device was added as an alternative device to the first device identified.

Based on the descriptions of the identified hardware the following categories (see 4) were extracted for each facet (type of experience, delivery technology and back-end technology):

Table 4: Hardware—Related Categories.

| Taxonomy facets | Categories |
| --- | --- |
| Type of Experience | Virtual Reality |
|  | Augmented Reality |
|  | Display |
| Delivery Technology | Head-Mounted Display |
|  | Screen |
| Back-End Technology | Stationary |
|  | Mobile |
|  | Cloud |

## 6.2 Software Systematic Mapping Study

### 6.2.1 Study planning

The aim of this systematic mapping study was to discover non-overlapping categories which would allow grouping of the VR/AR applications used in education. The relevant research question is "What VR/AR applications are being currently used for education purposes?".

The keywords and alternative words were derived from the research question are summarized in 5.

Table 5: Software—Included Terms.

| Main terms | Alternative terms | Additional terms |
| --- | --- | --- |
| Virtual Reality, Virtual Worlds, Mixed Reality, Augmented Reality, Gamification, Virtual Training, Virtual Seminar, E-Learning, Cinematography | VR, XR, AR, 3D Cinema | Classroom, Teaching, Applications, Software, Telepresence, Stereoscopic |

A search was performed and appropriate papers were selected according to the selection strategy. Educational applications which authors described as virtual or augmented reality (or the related terms) were compiled into a list.

### 6.2.2 Study results

Ten distinct applications were identified with the following observations:

- Four used a general purpose controller, one used both a general purpose controller and user gaze functionality, three used a special controller and two used no controller at all.

- Three used a dialog system to interact with the user, three used intelligent agents and four did not use any interaction with the user.

- Seven were created using 3D modeling and three using cinematography.

- One used reward-based gamification, one employed principles of serious games and eight did not use any gamification.

Thus the categories (see 6) were extracted from the descriptions of the devices for each facet (type of experience, delivery technology and back-end technology):

| Top Hierarchy | Design Requirements | | Physical Requirements | | Interface Requirements | | Implementation Requirements | |
|---|---|---|---|---|---|---|---|---|
| Facets | Educational Purpose | Type of Experience | Delivery Technology | Back-End Technology | User Interaction | System Interaction | Production Technology | Type of Gamification |
| Categories | Training<br>Teaching<br>Observing | Virtual Reality<br>Augmented Reality<br>Display | Screen<br>Head Mounted Display | Stationary<br>Mobile<br>Cloud | General Purpose Controller<br>User Tracking<br>Specialized Controller<br>No Interaction | Dialog System<br>Intelligent Agents<br>No Interaction | Modeling<br>Cinematography | Reward Based<br>Serious Games<br>No Gamification |

Figure 2: Taxonomy overview

Table 6: Software—Related Categories.

| Taxonomy facets | Categories |
|---|---|
| User Interaction | General Purpose Controller<br>User Tracking<br>Specialized Controller<br>No Controller |
| System Interaction | Dialog System<br>Intelligent Agents<br>No Interaction |
| Production Technology | 3D Modeling<br>Cinematography |
| Type of Gamification | Reward-Based Gamification<br>Serious Gamification<br>No Gamification |

# 7 Taxonomy Overview

An overview of the taxonomy is given on 2. Its structure reflects the decision process when designing a new application or deciding the suitability of using a specific application in a teaching and learning situation. The taxonomy employs successive decomposition (top-down analysis), which allows for the analysis to focus on specific functions and to avoid confusion from focusing on less important details. Specifically, attention is given to:

1. *Design Requirements*

    These are decided first by the educator when designing or choosing the application. They define the educational purpose of the new application and the type of experience desired.

2. *Physical Requirements*

    These specify what delivery technology and back-end technology is going to be used and is largely determined by the price of the devices (such as screens being cheaper than HMDs). Technology requirements also affect the content requirements.

3. *Interface Requirements*

    These are decided based on the physical requirements and are affected by the design requirements. For example, if the delivery technology is HMD and the back-end technology is mobile, then the complexity of 3D models has to be lowered.

4. *Implementation Requirements*

   These define how the experience (targeting a specific physical device with a defined interface) is created by the developers.

Each of these top hierarchy points contain two facets and each facet contains some number of categories. Each individual category of the taxonomy can be used to describe diverse devices or principles. The description of the VR/AR experience becomes more specific with the selection of combinations of these categories.

## 7.1 Classification of VR/AR by Educational Purpose

The educational purpose of the application should be clear to the educator before implementing it in the learning process, and to the student before using it. An ability to understand the purpose of the application by the user increases its perceived usefulness and motivation for use (Beke Hen, 2019). The three non-overlapping categories in the proposed taxonomy are training, teaching and observing.

### 7.1.1 Training

For training purposes, the goal of the application is to convey information about how to use a specific real device. It is very specific in purpose, with the focus on training for equipment, machine or process operation rather than the understanding of the underlying principles of design. Examples include medical training such as dentistry (Roy, Bakr, and George, 2017), laparoscopy (Wynn, Lykoudis, and Berlingieri, 2017) and ophthalmoscopy (A. S. Wilson et al., 2017), or use in the military (Annetta, 2010). Implications for consideration include:

- *Implications for the development:* precise in function and/or visual appearance.
- *Implications for the educator:* tasks aiming to help students internalize smaller pieces of information, skill, behavior, etc. Effective internalization might be achieved through repetition.

### 7.1.2 Teaching

For teaching purposes, the goal is to prepare the student to remember, recollect and understand knowledge in a general situation. The student is being exposed to theory and underlying principles, and such knowledge is expected to be transferable to other situations and environments. Currently, the number of examples of successful teaching applications is relatively low, as often their development is challenging (Holmes, 2007) with success relying on effective scaffolding (Coleman, 1998) as well as effective integration of assessment and feedback. Examples exist where no advantage in using immersive VR was found (e.g., (Moro et al., 2017)). However, some examples of effective applications exist in areas of language teaching (Shih and Yang, 2008), general lab work Ren et al. 2015; Gardner et al. 2019 and agriculture (Laurel, 2016). Implications for consideration include:

- *Implications for the development:* the virtual environment does not need to be precise.
- *Implications for the educator:* the tasks might be larger in size and with less repetition.

### 7.1.3 Observing

For observing purposes, the primary goal is to motivate the students, to show or convey information without the need for retaining or understanding it. Examples of such applications can be seen in the form of historical recreations of sights (Barreau et al., 2015) or artifacts (Nicolas et al., 2015) (the latter allowing for example shared analysis and research of objects between universities). When the student is confronted with information transcending previous knowledge schemas, the student can feel complex emotion known as "awe" which includes feelings of astonishment, wonder and connectedness (Keltner and Haidt, 2010). Implications for consideration include:

- *Implications for the development:* considerable effort must focus on the visual appearance and awe evoked by the virtual environment and/or objects.
- *Implications for the educator:* students will not remember details. Students should feel motivated to learn more about the subject.

## 7.2 Classification by Experience

Whilst some authors differentiate between VR and AR experiences based on the level of isolation of the user from the outside environment (Schmalstieg and Hollerer, 2016; Velev and Zlateva, 2017), this taxonomy focuses on whether the user is presented with objects and environments that are not around the user at the given moment (such as providing an experience of being in a chemical laboratory (Gardner et al., 2019) or executing microsurgery on a patient (Wynn, Lykoudis, and Berlingieri, 2017)) or whether some information is added on top of the real world that is surrounding the user. Also, this taxonomy extends and aims to allow description of technologies such as *virtual worlds*, given the commonplace confusion amongst educators and researchers as to whether these are a part of the *virtual reality* space or not (e.g., (Warburton, 2009)). The proposed taxonomy identifies three distinct user experiences, Virtual Reality experience, Augmented Reality experience and Display experience.

### 7.2.1 Virtual Reality experience

The VR experience focuses on presenting the user with objects or entire environments which do not exist around them at the given time. It could be visual (displaying environment which is not real), or presenting the user with an imitation (a virtual object), simulating behavior of its real counterpart (such as laparoscopic training kit for "virtual" surgery). The user might be completely isolated (as proposed by (Schmalstieg and Hollerer, 2016)) or immersed in a way that allows he user to focus on the virtual "life like" object of the experience, creating a strong illusion of presence (Milleville-Pennel and Charron, 2015). The VR experience can be delivered using various devices, such as a HMD (Fuchs, 2019), but also include driving simulators, a set of displays positioned to surround the user (CAVE systems (Cruz-Neira, Sandin, and DeFanti, Sep. 1, 1993)) or a physical machine which provides greater haptic realism, such asan endoscope analogue interfaced to a computer screen (M. S. Wilson et al., 1997). Implications for consideration include:

- *Implications for the development:* the complete environment must be designed or otherwise digitized.
- *Implications for the educator:* student immersion must facilitate the intended educational goal. Student's will not be able to make notes.

### 7.2.2 Augmented Reality experience

AR systems combine (overlay) virtual content (e.g., generated through a model, animation or video recording) with real-world imagery (Klopfer and Squire, 2007). Therefore, AR adds information to the reality surrounding the user. This occurs in real time as the user engages with the system, with aspects of the surroundings or other real-world objects registered in 3D (Azuma, 1997). Two examples that demonstrate the relevance of such characteristics are described below:

1. *AR annotation applications* are a common type of AR applications that display text and other symbols on real-world objects (Santos et al., 2014). These applications stand out as these symbols, such as labels, arrows and lines are not 3D objects, but need to be rendered in 3D space by the application in real time, whilst scanning the environment using image recognition to determine appropriate position for these labels.

2. *Vision-haptic visualization applications* are embodied interactions with virtual content (Santos et al., 2014). Examples include an application which allows the user to view a 3D model on a marker which they can manipulate using their hands (Martın-Gutiérrez et al., 2010). Educational seffectivity of these applications is questionable as the findings show no improvement in learning outcomes (Martın-Gutiérrez et al., 2010).

In terms of delivering the AR experience, it can be achieved using a screen (monoscopic) or HMD (stereoscopic). HMD for AR can be described as video see-through closed-view HMD(Azuma, 1997), that means the image is captured with a camera, blend together with a 3D generated image and displayed through an HMD resembling the VR headset. Another possible way an AR HMD can be constructed is by using see-through panels with a 3D image projected on them, which means that in the case of a loss of power the user can still see the real world, which has implications on safety, especially when used in industry contexts. Implications for consideration include:

- *Implications for the development:* technology allowing the blending of the virtual and real must be used.
- *Implications for the educator:* access to a specific place or objects might be required. Real world augmentation must facilitate the intended educational goal.

### 7.2.3 Display experience

The content is shown on a display or monitor and might be immersive (e.g., 3D generated world in which the user can freely walk) but it is not presented in a way that would create a strong illusion of presence (Milleville-Pennel and Charron, 2015). A display experience offers some advantages over an otherwise more immersive VR, such as the relative clarity of text and general reading experience compared to the capabilities of modern HMDs, and is much less prone to causing user dizziness, headache or eyestrain (Penczek et al., 2015). In terms of the educational advantages, the user can easily make notes.

## 7.3 Classification by Delivery Technology

The user experience may be delivered through two distinct hardware devices: screen and HMD. Both types of devices can provide monoscopic as well as stereoscopic vision, although, 2D is more commonly delivered using the screen (as 3D screens are not as common) and 3D is commonly delivered using HMD (as displaying 2D in HMD is not economical).

### 7.3.1 Head-Mounted Display

The fundamental idea behind the stereoscopic head-mounted display (HMD) is to present the user with a perspective image which changes as the user moves (Sutherland, Dec. 9–11, 1968). HMD allows stereoscopic vision in VR (Fuchs, 2019) as well as in AR (Schmalstieg and Hollerer, 2016). Stereoscopic vision arises when two views of the same scene with binocular disparity are presented to each eye. The effect depends on binocular fusion in order to yield perception of depth (Hale and Stanney, 2014). In both cases, user's hands are free (i.e., the user does not have to hold the device in their hands). Adverse effects of low refresh rates of VR HMDs have been observed since at least 1993, refresh rates of at least 60Hz are recommended and can be achieved by carefully limiting the complexity of the scenes (Ware, Arthur, and Booth, 1993). By its nature, the HMD is not a shared experience (as not shared in real world, the experience can still connect multiple users if the headset is connected with other devices through a computer network), therefore if the educator's goal is for the students to cooperate it is unlikely to be the right choice. In the past the developers could choose to use monoscopic vision in order to reduce the need for heavy graphics processing (Youngblut, 1998). Implications for consideration include:

- *Implications for the development:* a suitable programming interface for the given HMD needs to be used.

- *Implications for the educator:* need to have HMDs for the students. In case of a 6dof (degrees of freedom) experience, suitable space is needed. Interference might occur if more than one headset is used in a room.

### 7.3.2 Screen

The user uses a stationery (e.g., desktop computer) or hand-held (e.g., tablet) device (Pensieri and Pennacchini, 2014), or may in fact be surrounded by the screen as in the case of the CAVE system (Bruder et al., 2016). In either case, the screen does not allow 3D vision, with the exception of 3DTV. A common computer screen is an important part of some VR or AR implementations, for example a simulator of medical operation uses a display to show image from a camera of an endoscope. This, together with a haptic endoscope simulator creates a VR simulation of endoscope operation (Miki et al., 2016). One implication for the educator is that a stereoscopic view surrounding the student cannot be achieved.

## 7.4 Classification by Back-End Technology

Backend technology is the technology that makes it possible for VR/AR to function which is not in direct contact with the user.

### 7.4.1 Stationary

A stationary technology is immovable (or movable with great difficulty) or is not detachable from another VR/AR unrelated device (e.g., in the case of a car HUD). Higher computational power is achievable, leading to better graphics quality, more precise physics calculations and larger worlds which subsequently leads to higher level of immersion. Implications for consideration include:

- *Implications for the development:* more objects can appear on the screen and with higher quality. Development might be faster due to less concern over the performance.

- *Implications for the educator:* the student cannot easily move outside of the range of the stationary part (e.g., desktop computer). Manipulation with the device might be more difficult.

### 7.4.2 Mobile

A mobile technology is easily movable and the experience is generated directly on a device built to be carried around or worn. Mobile devices have lower performance but higher usability (no wires). Both VR and AR applications are more cost effective if developed for mobile technology; affordability makes it possible for this technology to be used on a large scale (Fuller and Joynes, 2014). Implications for consideration include:

- *Implications for the development:* additional optimization is needed which might slow down the development.
- *Implications for the educator:* the student can move around. Manipulation with the device might be easier.

### 7.4.3 Cloud

In cloud technology the main computation process providing the experience (such as graphics) is generated on a remote server, which is not part of the application (the user does not have a direct access to it). The delivery technology used for this experience is connected with a high speed wireless connection (e.g., 5G) to a common network in order to access the server. Implications for consideration include:

- *Implications for the development:* more objects can appear on the screen and with higher quality. Development might be faster due to less concern over the performance.
- *Implications for the educator:* the student can move around. Manipulation with the device might be easier. Access to the server must be maintained.

## 7.5 User Interaction

In order for the VR/AR application to be interactive, the user has to have some method of sending signals to the system. This typically involve tracked controllers (e.g., gloves or sticks), or if these are not available or desirable, a simpler application control can be employed. In some cases, no input, and therefore no user interaction, may be acceptable.

### 7.5.1 General-Purpose Controllers

The application uses general purpose controllers to receive input from the user. This type of controller is a default part of the device delivering the experience and is universal, allowing control of various applications supported by the technology. HMD delivery devices and their controllers are tracked in real-world 3D space and then projected into the virtual world. A typical general-purpose controller in case of a display experience can be for example computer mouse and keyboard, or a touch based display where the mechanism of the controller is part of the display (i.e., resistive or capacitive layer on top of the screen). Implications for consideration include:

- *Implications for the development:* easier to implement.
- *Implications for the educator:* muscle memory can not be trained (unless training focuses on the use of the general-purpose controllers, for example touch-typing training application). Interference between controllers might occur if more than one set is used.

### 7.5.2 User Tracking

The user-tracking system directly scans the position of parts of the user's body, such as eyes, hands or entire body. This includes controlling the application through gaze, hands or full body scanning.

### 7.5.3 Special Controllers

A special controller is a controller for input which cannot be replicated using a general purpose controller or by user tracking. These controllers are common in medical training and can include for example virtual endoscopes (Baier et al., 2016), simulators of dental procedures (Roy, Bakr, and George, 2017) or ophthalmoscopes (A. S. Wilson et al., 2017). Such controllers can also be simpler, such as turning wheels for drivers (Milleville-Pennel and Charron, 2015). Such controller can employ actuation feedback force, a technology that has been used since at least 1965 in flight training systems (Sutherland, 1965). Implications for consideration include:

- *Implications for the development:* harder to implement.
- *Implications for the educator:* allows muscle memory training but the controller must accurately represent the real object (see training purpose).

### 7.5.4 No Interaction

The user cannot send input to the VR/AR application and therefore the application cannot be interactive. This can be the case for applications based on cinematography, either as display experience or even a VR experience (e.g., 360 panoramic video).

## 7.6 System Interaction

System interaction categorizes how the VR/AR application communicates with the user information related to the educational goal. Three categories were identified: dialog system, intelligent agents, and no system interaction. Focus on educational properties of the application is important in order to define the system interaction information correctly. For example, where an application contains a dialog option to simply close the application it does not mean it uses a "dialog system". The application is classified as "dialog system" if dialogs are used to achieve the educational goals (e.g., through information panels).

### 7.6.1 Dialog Systems

According to explanation-based constructivist theories of learning, learning is more effective and deeper when the learner actively generates explanations rather than being merely presented with information (Aleven and Koedinger, 2002). This theory is being used by dialog systems, which ask the student to provide explanations of the educational context by means of menus or direct textual input. Effectiveness of learning is reported to be higher when the student is asked to answer questions via direct textual input (Aleven, Ogan, et al., 2004). Dialog systems are successfully used in non-VR/AR educational related applications (Graesser et al., 2005) but they are not as easily employed in such form in VR/AR because it is harder to implement an effective method of input, especially when the experience is delivered via HMD (Bowman, Rhoton, and Pinho, 2002). Implications for consideration include:

- *Implications for the development:* easier to implement.
- *Implications for the educator:* does not allow for complex interaction between the student and the application. Might break immersion.

### 7.6.2 Intelligent Agents

Intelligent agents are more sophisticated than dialog systems and interact with the user in a more complex way than just textual or audio information. The user can see their representation as an avatar which can move in the virtual space and operate objects in the virtual world (Rickel and Johnson, May 1, 1998), which adds life to the virtual world and improves immersion of the VR application (Ochs, Sabouret, and Corruble, 2010). Intelligent agents can have the same effectiveness as human tutoring (VanLehn, 2011). A lack of intelligent agents was identified as one of the problems of sustaining user immersion and interest in educational VR applications (Champion, 2015) and a number of authors planned to include such intelligent agents in future work (e.g., (Dede, Salzman, and Loftin, Mar. 30–Apr. 3, 1996) and (Barreau et al., 2015)). Implications for consideration include:

- *Implications for the development:* requires implementing non-player characters (NPCs). Scripted audio recordings might have to be recorded.
- *Implications for the educator:* complex interaction can be simulated. Does not break immersion.

### 7.6.3 No Interaction

In cases where the explanation of use or function are provided by other sources, the system may provide no interaction, for example a driving simulator. One implication for the educator is that the student needs to be informed or trained separately on how to achieve the goals.

## 7.7 Classification by Production Technology

In this taxonomy, it has been identified that it is possible to produce VR/AR experiences using modeling (e.g., computer generation, 3D modeling) or cinematography (e.g., stereoscopic 360 video recording).

### 7.7.1 Modeling

Modeling, 3D modeling and generated computer graphics is the most common approach to develop computer games and by extension serious games for education. 3D models can be designed using tools, such as Blender or using photogrammetry using 3D scanners (Nicolas et al., 2015). Implications for consideration include:

- *Implications for the development:* specific 3D modeling skills and knowledge of 3D techniques used to build 3D environments is required. 3D scanner is necessary if 3D scanning is employed. Can be time consuming.

- *Implications for the educator:* ability to display the object from any angle.

### 7.7.2 Cinematography

In terms of cinematography, the footage is filmed with a specific field of view, most commonly around 60 degrees. For use in VR it is usually between 180 – 360 degrees, which then affects how much the user is surrounded by the image. The footage may also be filmed stereoscopically, that is to provide the illusion of 3D. Implications for consideration include:

- *Implications for the development:* specific filmography skills and techniques are required. Use of suitable camera is required. Time consumption does not scale up with complexity of the scene.

- *Implications for the educator:* can record scenes impossible or very difficult to 3D model.

## 7.8 Classification by Type of Gamification

Gamification describes game-inspired techniques to engage students within the learning and interaction process. The purpose of gamification is to increase student motivation for learning or skills development. It is possible to either embed game elements into the learning environment (Monterrat et al., Jun. 21-25, 2015) or to integrate educational content into a game (Lameras et al., 2016). The latter are also referred to as serious games.

### 7.8.1 Reward-Based Gamification

Adding elements such as leaderboards, badges and achievements to the learning content in order to motivate students to progress through it. This on the whole may be seen as extrinsic motivators for the learning application. Evidence for efficiency and positive results of gamification has been shown in studies in the UK and US (Lameras et al., 2016). Gamification gives students rewards, motivating progress through the course; the rewards are usually intangible, such as the feeling of victory (Bodnar et al., 2015). Implications for consideration include:

- *Implications for the development:* relatively easy to implement

- *Implications for the educator:* possible lower efficiency to improve motivation. Easily quantified.

### 7.8.2 Serious Games

Serious games use game elements to increase students' internal motivation by adding educational content to the game. Examples of serious games include a computer-based game with a goal to teach cooperation between players (Dunwell et al., Apr. 24–May 1, 2014). A role-playing board game addressing water management and irrigation around Markala dam in Mali is another such example (Hertzog et al., 2014). Serious games can also provide stealth learning, for example a situation when the player is achieving the intended learning outcomes without realizing it (Annetta, 2010). Implications for consideration include:

Table 7: Experience and Delivery Technology—Devices.

|  | Screen | Head-Mounted Display |
|---|---|---|
| Display | Computer monitor, Tablet, Phone, 3DTV, Looking Glass | Google Glass, Scorpion |
| Virtual Reality | Screen-based driving simulator, CAVE, MIST VR | HTC Vive, Oculus Quest/Go |
| Augmented Reality | Tablet based AR, Heads-up display (HUD) | Meta, Microsoft HoloLens |

- *Implications for the development:* relatively hard to implement. Requires close cooperation with the educator. Might require complex interaction with the student.

- *Implications for the educator:* develops long-term motivation.

### 7.8.3 No Gamification

No gamification is present in VR/AR application if it does not focus on supporting students' motivation by presenting them with gamification elements. In many cases (e.g., driving simulator or laparoscopy simulator) gamification is not necessary and might be detrimental to the learning process if not implemented seamlessly (Vogel et al., 2006).

## 8 Use of the Taxonomy

The taxonomy becomes increasingly specific with the number of categories used to describe a VR/AR experience. For example, {Virtual Reality, Display, Screen, Mobile} and {Virtual Reality, Display, Screen, Stationary} distinguish between a tablet sized 3D environment and a large CAVE system. Both these systems have commonalities, such as having a potential to be a classroom shared experience, this comes from the fact that both have the same root {Virtual Reality, Display, Screen}. Further specification gives a very specific idea about the hardware required, its cost, potential for immersion, how the content will be consumed, scalability, general educational goal, etc. These properties, including mobility and immersion emerge from combinations of our taxonomy's categories. Thus the design implications of the taxonomy can be extracted by considering the intersections between the taxonomy facets, such as affordability, level of interaction, etc. (see 2). This section demonstrates how VR/AR devices are categorized using the taxonomy and analyses these intersections. The following specific design implications are explored: how the choice of an experience and delivery technology affects the hardware selection; and how the taxonomy facets affect the teaching method and the cost of implementing the application. In addition, two case studies where the taxonomy was used in practice to aid the development process of VR/AR applications are presented.

### 8.1 Experience and Delivery Technology Implications on Hardware

Combination of "Type of Experience" described in the previous subsection, and "Delivery Technology"' determines which device to implement for the desired experience (see 7).

Specific points of interest for the experience and delivery technology combinations are given below:

1. *Display and Screen (Computer monitor, Tablet, Phone):* the Display experience can be provided by a large number of devices such as a computer monitor.

2. *Display and Head-Mounted Display (Google Glass, Scorpion):* this category of devices is small as the practicality of having an HMD device which does not provide VR/AR experience is limited. An example of such a device was developed by Google, known as the Google Glass, actually offers a screen experience inside an HMD. The reason why Google Glass is not an AR experience relates to the type of information Google Glass is showing. Since it is not showing information related to what is in front of the user, nor is registered in 3D (Google glass does not have the necessary sensors for that), it is basically a screen showing information similar to what the smart phone does (but strapped onto the user's head). Therefore, it should not be considered an AR device.

3. *Virtual Reality and Screen (Screen based driving simulator, CAVE, MIST VR):* devices which use screens (usually multiple) in order to produce a virtual environment, often together with auditory and/or sensory stimulation. Examples of these devices are screen based driving simulators inside of which the person sits and is surrounded by multiple screens. Even with the use of just a few screens, such a simulator can feel immersive and create a strong illusion of presence, comparable with the real-world object (Milleville-Pennel and Charron, 2015). Another example is the CAVE system (Cruz-Neira, Sandin, and DeFanti, Sep. 1, 1993).

4. *Virtual Reality and Head-Mounted Display (HTC Vive, Oculus Quest/Rift):* there has been a significant amount of development regarding HMDs in recent years as these became widely available to users. A considerable disadvantage in terms of their educational use is the subsequent inability or great difficulty for users to make notes in real life as well as within the virtual world (Poupyrev, Tomokazu, and Weghorst, 1998).

5. *Augmented Reality and Screen (Tablet based AR, Heads-up display (HUD)):* HUDs, information screens projected in front of the users (e.g., drivers, pilots), originally developed for military flight simulation but with great potential in the automotive industry (Chu, Brewer, and Joseph, 2008) where the information is projected on the windshield (Schneider et al., Sep. 22–25, 2019) using a dedicated device or a mobile phone (Prabhakar et al., 2019). Whilst the technology has great potential, increased cognitive demands on the driver need to be considered (Wolffsohn et al., 1998) as well as the position of the projected image (You et al., Sep. 24–27, 2017) in order to avoid cognitive tunneling (Foyle, Dowell, and Hooey, 2001).

## 8.2 Implications of Selected Facets on Teaching Method

Various teaching methods (such as direct instruction, interactive whole class teaching, student-centered approaches (Westwood, 2008)) can be supported or even completely managed by the use of AR/VR applications (e.g., a dancing tutor (Chan et al., 2011)). Each facet of the proposed taxonomy, except production technology, has impact on a teaching method employed by the educator:

1. *Type of Experience*: AR allows for more interaction between the participants as by definition the user can see their surroundings. VR experiences might (but do not have to) isolate the user, especially if the chosen delivery technology is an HMD as they are isolating the user (in some cases almost completely). This has implications on the educational style used.

2. *Delivery Technology:* HMDs tend to isolate the user more than the screen delivery technology; the latter is more suitable for groups of people.

3. *Back-End Technology:* mobile and cloud back-end technology might be more suitable for larger groups due to better affordability (see more in 8.4).

4. *User Interaction:* user tracking and general-purpose controllers might be more suitable for larger groups due to better affordability (see more in 8.4).

5. *System Interaction:* dialog systems and intelligent agents are more suitable for self-study as they can guide the student. More teacher-led approaches will be required if no system interaction is provided.

6. *Type of Gamification:* reward-based gamification might be more suitable for classes as it provides possibility for competition between students. Serious games might be more suitable for self-study as it offers guidance for the student and might be less suitable for classes unless specifically designed for larger numbers of participants.

## 8.3 Implications of Experience and Delivery Technology Facets Intersections on Teaching Methods

The implications on teaching method for individual intersections between experience and delivery technology are summarized below:

1. *Display and Screen (Computer monitor, Tablet, Phone):* suitable for a large classroom as well as individual tutoring, but might not be engaging enough for self study (Clow, Apr. 8–13, 2013). Students are not fully immersed in the experience but might cooperate easily and shift their focus from the experience to their surroundings and peers as necessary. They can easily make notes.

2. *Display and Head-Mounted Display (Google Glass, Scorpion):* the class size is limited by the number of devices available.

3. *Virtual Reality and Screen (Screen based driving simulator, CAVE, MIST VR):* some technologies (eg CAVE, LAP Mentor) allow for participation of multiple students in one experience (usually not a large class). Students need to have access to the delivery technology for self study.

4. *Virtual Reality and Head-Mounted Display (HTC Vive, Oculus Quest/Rift):* students cannot see each other, a lot of space is required if the experience requires them to walk. Class size is limited by the number of devices available, but also interference might occur if multiple devices are being used next to each other. Likewise, the classroom may be difficult to manage in the case of complex applications.

5. *Augmented Reality and Screen (Tablet based AR, Heads-up display (HUD)):* students hold a mobile device or are positioned in front of a (large) screen, thus readily scalable to large groupsbut may be limited by the environment that is being augmented.

6. *Augmented Reality and Head-Mounted Display (HoloLens, Google Glass):* students wear headsets but can see each other; headsets may be used to navigate the user around the surroundings they are in.

## 8.4 Cost Analysis

The proposed taxonomy can help with cost analysis of a VR/AR application which is being developed or assessed for purchase by educators and can serve as a basis for return on investment metrics both financial (e.g., cost reductions) and non-financial (e.g., saved time, improved satisfaction). Some of the cost implications of the taxonomy facets and their categories include:

1. *Type of Experience:* there might be an increase of costs in the use of AR because it overlays information on top of the real-world object or environment. The cost increases if access to the object or object itself needs to be purchased. The display experience is the simplest, cheapest and most traditional experiencebut offers lower immersion.

2. *Delivery Technology:* immersive screen experiences (such as CAVE) can significantly increase the cost (Hilfert and König, 2016). Cloud based delivery can decrease the cost of the users' devices as these can be technologically simpler but introduces a new cost for the server (renting or running/maintaining).

3. *Back-End Technology:* stationary VR/AR experiences need to be connected to a device which provides or allows for correct function of the experience whereas the mobile devices and devices connecting to the cloud are more light weight, albeit less capable. For this reason mobile devices and devices connecting to the cloud server seem to be more cost effective than stationary experiences (Ford et al., 2018).

4. *User Interaction:* when designing a new VR-HMD/AR experience, the way the user interacts with the system is one of the cost driving factors. General purpose controllers and user tracking is more affordable than special controllers, some of which need to be calibrated in order to provide successful learning (Våpenstad et al., 2017).

5. *System Interaction:* intelligent agents require more programming effort (such as use of behavior trees (Colledanchise and Ögren, 2018)).

6. *Production Technology:* modeling and 3D modeling is time consuming and expensive. Cinematography and even stereoscopic cinematography might be cheaper and allow for capturing more detail than what would be achieved with 3D modeling, but offers less flexibility as the scene can not be easily changed after the recording has been finished. VR applications produced with cinematography are also less hardware demanding, further reducing the cost of the delivery technology (Brown and Green, 2016).

7. *Type of Gamification:* reward-based gamification can be added to already existing educational materials and therefore be more cost effective, although gamification is proving to be less effective than serious games. Serious games can be expensive to design as making a connection between educational material and play experience must be seamless otherwise the system might inhibit the ability and motivation of users to learn new material (Vogel et al., 2006). Therefore, gamification might not be necessary at all, and is typically not present in examples of applications aiming at training educational purpose (e.g., driving simulator) which might decrease cost due to simpler design.

## 8.5 Case Study: VR Chemical Plant

The VR Chemical Plant was devised as a platform on top of which varied educational content could be created, such as assignments and tasks with the ability for the content to provide training, teaching and observing (each in separate tasks). A specific task was created aiming to help students learn the location of important plant equipment and instrumentation. The educational goal was based on module learning outcomes, for example locating key valves on the plant, and falls within the training category of the education purpose facet. This implies that the visual appearance and location of the valves must precisely match the real-world plant. Also, from the educator's perspective, the task given to the students should be shorter and specific: in this case the task is to find a valve which controls a specific function on the plant (chosen randomly from 21 valves).

Given the aim of resolving logistical difficulties associated with student access and use of an actual chemical plant and the related safety considerations, the VR experience was chosen over AR to offer an immersive experience. This means that the entire plant needs to be digitized. The use of VR is justified by the fact that the aim is to teach spatial knowledge (position of the valves). In connection with the previous point and in order to increase immersion of the students, use of HMD was decided and the compatible SteamVR programming interface used together with the game engine Unity3D. A room was designated to be used for the purpose of testing the application.

Given that student movement was restricted to the designated room and the requirements of high visual quality, a stationary, desktop computer was used to provide the processing power. This allowed the development process to be faster. The student needs to be able to move around the VR Plant in order to locate the valves, therefore some controllers must be implemented. However, the training on the VR Plant does not involve muscle memory, therefore default general purpose controllers could be used. In order to provide complex interaction which would not break student immersion, several NPCs were designed and dubbed by volunteering students. The interaction could have been realized as a much simpler dialogue system; this step increased the time required to build the simulation. The task required the student to be able to walk around in an attempt to find the correct valve. For that reason, it was decided to use 3D modeling in combination with 3D scanning to create the 3D model of the plant. Since it was decided to develop NPCs which allow for complex interaction, a serious games approach was chosen. Given the choice of an HMD for VR, HTC Vive was selected as a delivery technology. Only one student uses the application at any given time and therefore no interference can occur.

As shown above, choices associated with each facet had to be made and their implications considered before development of the application begun.

## 8.6 Case Study: Walking Dog

The AR application Walking Dog was developed to help veterinary medicine students recognize a dog's limp through observation of the dog walking. The simulation involves 3D model of a dog walking in a circle; students can observe the dog on an iPad screen seamlessly embedded in the real world. The educational content does not focus on teaching the students principles of recognition of lameness (e.g., how the weight is shifted or how the head and neck position is changing), this information needs to be presented to the students during the appropriate lectures. Students then use this information to train themselves in recognition of lameness and identification of its location and its severity. The educational purpose is therefore 'training'. The definition of this facet helps the educator understand that the students will be making multiple attempts in recognition of lameness. For this reason it was decided that the application should include testing functionality, during which the dog exhibits limping on a random leg and with random severity. Students test themselves in its recognition.

The Augmented Reality experience was selected for multiple reasons. High immersion is not necessary for successful training and would likely prevent cooperation between students (unless an expensive CAVE system was used). For this reason, it was chosen to use screen, rather than HMD. These choices allowed for use of a relatively cheap mobile device which makingit readily possible for classroom use.

Students need to be able to move around as they follow the dog walking. The activity is performed as part of a class in a large classroom. Students receive all the information necessary to make a correct diagnosis as part of their class, for this reason there is no complex interaction required and the application relies on dialog systems guiding the student through the testing process. A relatively simple 3D model of a dog was used together with bone animation prerecorded using motion capture. No gamification was programmed in the application due to constraints on development time, this can be supplemented by letting students compete during the class room activity.

Again, choices associated with each facet had to be made and their implications considered before development of the application begun.

# 9 Conclusion

The interest in VR/AR generated by educators in recent years reflects the speed with which the technology is being developed and how affordable it has become. The number of applications of VR/AR is large and varied. To make the field more understandable and to make the meaning behind technical language clear, a need for a new taxonomy has been identified. A clear methodology was outlined and facets of the new taxonomy established. Categories of the "educational purpose" facet have been determined based on the existing Bloom's taxonomy of educational objectives. A systematic mapping study of hardware used to facilitate VR/AR has been performed and categories of taxonomy facets of Type of Experience, Delivery Technology, and Back-End Technology were identified. A second mapping study aiming to map the VR/AR applications was also conducted with the resulting facet categories of User Interaction, System Interaction, Production Technology and Type of Gamification. Implications of the facets and some of their combinations on teaching method, suitable hardware and cost have been described. Thus, the proposed taxonomy provides clarity on the classification of VR/AR and should facilitate more effective communication between educators, developers, and researchers. Future work will include the establishment of a framework for development of these experiences based on the presented taxonomy, thus allowing the development process to be more structured and manageable.

# References

Aleven, Vincent and Kenneth R Koedinger (2002). "An effective metacognitive strategy: learning by doing and explaining with a computer-based Cognitive Tutor". In: *Cogn. Sci.* 26.2. doi: 10.1016/S0364-0213(02)00061-7, pp. 147–179.

Aleven, Vincent, Amy Ogan, et al. (2004). "Evaluating the Effectiveness of a Tutorial Dialogue System for Self-Explanation". In: *Intelligent Tutoring Systems*. doi: 10.1007/978-3-540-30139-4_42. Berlin, Heidelberg: Springer, pp. 443–454. ISBN: 978-3-540-22948-3.

Annetta, Leonard A (2010). "The "I's" have it: A framework for serious educational game design". In: *Rev. of General Psychol.* 14.2. doi: 10.1037/a0018985, pp. 105–112.

Azuma, Ronald T (1997). "A Survey of Augmented Reality". In: *Presence: Teleoperators and Virtual Environ.* doi: https://doi.org/10.1162/pres.1997.6.4.355, pp. 1–48.

Baier, Pablo A et al. (2016). "Simulator for Minimally Invasive Vascular Interventions: Hardware and Software". In: *Presence: Teleoperators and Virtual Environ.* 25.2. doi: 10.1162/PRES_a_00250, pp. 108–128.

Barreau, Jean-Baptiste et al. (2015). "An immersive virtual sailing on the 18th-century ship Le Boullongne". In: *Presence: Teleoperators and Virtual Environ.* 24.3. doi: 10.1162/PRES_a_00231, pp. 201–219.

Beke Hen, Libi (2019). "Exploring Surgeon's Acceptance of Virtual Reality Headset for Training". In: *Augmented Reality and Virtual Reality*. doi: 10.1007/978-3-030-06246-0_21. Cham: Springer International Publishing, pp. 291–304. ISBN: 978-3-030-06245-3.

Bloom, Benjamin S et al. (1956). *Taxonomy of Educational Objectives, Handbook I: Cognitive Domain*. doi: 10.4135/9781412958806.n446. London, U.K.: Longmans.

Bodnar, Cheryl A et al. (2015). "Engineers at Play: Games as Teaching Tools for Undergraduate Engineering Students". In: *J. of Eng. Educ.* 105.1. doi: 10.1002/jee.20106, pp. 147–200.

Bowman, Doug A, Christopher J Rhoton, and Marcio S Pinho (2002). "Text Input Techniques for Immersive Virtual Environments: An Empirical Comparison". In: *Proc. Human Factors and Ergonom. Society Ann. Meeting* 46.26. doi: 10.1177/154193120204602611, pp. 2154–2158.

Brereton, Pearl et al. (2007). "Lessons from applying the systematic literature review process within the software engineering domain". In: *J. of Syst. and Soft.* 80.4. doi: 10.1016/j.jss.2006.07.009, pp. 571–583.

Brown, Abbie and Tim Green (2016). "Virtual Reality: Low-Cost Tools and Resources for the Classroom". In: *TechTrends* 60.5. doi: 10.1007/s11528-016-0102-z, pp. 517–519.

Bruder, Gerd et al. (2016). "CAVE Size Matters: Effects of Screen Distance and Parallax on Distance Estimation in Large Immersive Display Setups". In: *Presence: Teleoperators and Virtual Environ.* 25.1. doi: 10.1162/PRES_a_00241, pp. 1–16.

Champion, Erik (2015). "Defining Cultural Agents for Virtual Heritage Environments". In: *Presence: Teleoperators and Virtual Environ.* 24.3. doi: 10.1162/PRES_a_00234, pp. 179–186.

Chan, Jacky C P et al. (2011). "A Virtual Reality Dance Training System Using Motion Capture Technology". In: *IEEE Trans. on Learning Technologies* 4.2. doi: 10.1109/TLT.2010.27, pp. 187–195.

Chirico, Alice et al. (2018). "Designing Awe in Virtual Reality: An Experimental Study". In: *Frontiers in Psychol.* 8. doi: 10.3389/fpsyg.2017.02351, pp. 105–14.

Chu, K, R Brewer, and S Joseph (2008). "Traffic and Navigation Support through an Automobile Heads Up Display (A-HUD)". In:

Clow, Doug (Apr. 8–13, 2013). "MOOCs and the Funnel of Participation". In: *Proc. 3rd Int. Conf. Learning Analytics & Knowledge (LAK'13)*. doi: https://doi.org/10.1145/2460296.2460332. Leuven, Belgium, pp. 185–189.

Coleman, Elaine B (1998). "Using Explanatory Knowledge During Collaborative Problem Solving in Science". In: *Journal of the Learning Sciences* 7.3-4. doi: 10.1080/10508406.1998.9672059, pp. 387–427.

Colledanchise, Michele and Petter Ögren (2018). *Behavior Trees in Robotics and AI: An Introduction*. doi: 10.1201/9780429489105. CRC Press. ISBN: 9780429489105.

Cruz-Neira, Carolina, Daniel J Sandin, and Thomas A DeFanti (Sep. 1, 1993). "Surround-screen projection-based virtual reality: the design and implementation of the CAVE". In: *Proc. 20th ann. conf. Computer graphics and interactive techniques (SIGGRAPH '93)*. doi: https://doi.org/10.1145/166117.166134. Anaheim, CA, USA, pp. 135–142.

Dede, C, M C Salzman, and Bowen Loftin (Mar. 30–Apr. 3, 1996). "ScienceSpace: Virtual realities for learning complex and abstract scientific concepts". In: *IEEE Proc. 1996 Virtual Reality Ann. Int. Symposium*. doi: 10.1109/VRAIS.1996.490534.

Douglas-Lenders, Rachel Claire, Peter Jeffrey Holland, and Belinda Allen (2017). "Building a better workforce: A case study in management simulations and experiential learning in the construction industry". In: *Educ. + Training* 59.1. doi: 10.1108/ET-10-2015-0095, pp. 2–14.

Duncan, Ishbel, Alan Miller, and Shangyi Jiang (2012). "A taxonomy of virtual worlds usage in education". In: *British Journal of Educational Technology* 43.6. doi: 10.1111/j.1467-8535.2011.01263.x, pp. 949–964.

Dunwell, Ian et al. (Apr. 24–May 1, 2014). "A game-based learning approach to road safety: the code of everand". In: *Proc. SIGCHI Conf. Human Factors in Computing Systems (CHI '14)*. doi: 10.1145/2556288.2557281. Toronto, Canada: ACM Press, pp. 3389–3398. ISBN: 9781450324731.

Eeles, Peter (2001). *Capturing architectural requirements*. Armonk, New York, NY, USA, IBM Corp., White Paper. URL: https://www.researchgate.net/publication/329760910_Capturing_Architectural_Requirements.

Farmani, Yasin and Robert J Teather (2020). "Evaluating discrete viewpoint control to reduce cybersickness in virtual reality". In: *Virtual Reality* 24. doi: 10.1007/s10055-020-00425-x, pp. 645–664.

Ford, Cameron G et al. (2018). "Assessing the feasibility of implementing low-cost virtual reality therapy during routine burn care". In: *Burns* 44.4. doi: 10.1016/j.burns.2017.11.020, pp. 886–895.

Foyle, D C, S R Dowell, and B L Hooey (2001). "Cognitive tunneling in head-up display (HUD) superimposed symbology: Effects of information location". In: *Proc. 11th Int. Symposium on Aviation Psychology*. Columbus, OH, USA.

Fuchs, Philippe (2019). *Virtual Reality Headsets — A Theoretical and Pragmatic Approach*. CRC Press. ISBN: 9780367888350.

Fuller, Richard and Viktoria Joynes (2014). "Should mobile learning be compulsory for preparing students for learning in the workplace?" In: *Brit. J. of Educ. Tech.* 46.1. doi: 10.1111/bjet.12134, pp. 153–158.

Gardner, Aaron et al. (2019). *Labster Virtual Lab Experiments: Basic Biochemistry*. Springer. ISBN: 978-3-662-58499-6.

Graesser, A C et al. (2005). "AutoTutor: An Intelligent Tutoring System With Mixed-Initiative Dialogue". In: *IEEE Trans. on Education* 48.4. doi: 10.1109/TE.2005.856149, pp. 612–618.

Hale, Kelly S and Kay M Stanney (2014). *Handbook of Virtual Environments*. Design, Implementation, and Applications, Second Edition. CRC Press. ISBN: 1466511842.

Hertzog, Thomas et al. (2014). "A role playing game to address future water management issues in a large irrigated system: Experience from Mali". In: *Agricultural Water Manage.* 137. doi: 10.1016/j.agwat.2014.02.003, pp. 1–14. ISSN: 0378-3774.

Hilfert, Thomas and Markus König (2016). "Low-cost virtual reality environment for engineering and construction". In: *Visualization in Eng.* 4.1. doi: 10.1186/s40327-015-0031-5, pp. 1–18.

Holmes, Jeffrey (2007). "Designing agents to support learning by explaining". In: *Comput. & Educ.* 48.4. doi: 10.1016/j.compedu.2005.02.007, pp. 523–547.

Hugues, Olivier, Philippe Fuchs, and Olivier Nannipieri (2011). "New Augmented Reality Taxonomy: Technologies and Features of Augmented Environment". In: *Handbook of Augmented Reality*. doi: 10.1007/978-1-4614-0064-6_2. B. Furht, Ed. New York, NY, USA: Springer, pp. 47–63. ISBN: 978-1-4614-0063-9.

Keltner, Dacher and Jonathan Haidt (2010). "Approaching awe, a moral, spiritual, and aesthetic emotion". In: *Cognition and Emotion* 17.2. doi: 10.1080/02699930302297, pp. 297–314.

Kitchenham, B and S Charters (2007). *Guidelines for performing Systematic Literature reviews in Software Engineering*. Tech. rep. EBSE 2007-001. School of Comput. Sci. & Math., Univ. Keele, U.K. and Dept. Comput. Sci., Univ. Durham, U.K.

Klopfer, Eric and Kurt Squire (2007). "Environmental Detectives—the development of an augmented reality platform for environmental simulations". In: *Educ. Technol. Res. and Develop.* 56.2. doi: 10.1007/s11423-007-9037-6, pp. 203–228.

Kwasnik, Barbara H (1999). "The Role of Classification in Knowledge Representation and Discovery". In: *Library Trends*.

Lameras, Petros et al. (2016). "Essential features of serious games design in higher education: Linking learning attributes to game mechanics". In: *Brit. J. of Educ. Tech.* 48.4. doi: 10.1111/bjet.12467, pp. 972–994.

Laurel, Brenda (2016). "AR and VR: Cultivating the Garden". In: *Presence: Teleoperators and Virtual Environ.* 25.3. doi: 10.1162/PRES_a_00267, pp. 253–266.

Leontyev, Aleksei Nikolaevich (1978). *Activity, Consciousness, and Personality*. Prentice Hall. ISBN: 978-0130035332.

Martın-Gutiérrez, Jorge et al. (2010). "Design and validation of an augmented book for spatial abilities development in engineering students". In: *Comput. and Graph.* 34.1. doi: 10.1016/j.cag.2009.11.003, pp. 77–91.

Mekni, M and Andre Lemieux (2014). "Augmented reality: Applications, challenges and future trends". In: *Applied comput. and applied comput. sci.*

Miki, Takehiro et al. (2016). "Development of a virtual reality training system for endoscope-assisted submandibular gland removal". In: *J. of Cranio-Maxillofacial Surgery* 44.11. doi: 10.1016/j.jcms.2016.08.018, pp. 1800–1805.

Mikropoulos, Tassos A and Antonis Natsis (2011). "Educational virtual environments: A ten-year review of empirical research (1999-2009)". In: *Comput. & Educ.* 56.3. doi: 10.1016/j.compedu.2010.10.020, pp. 769–780.

Milgram, Paul and Fumio Kishino (1994). "A Taxonomy of Mixed Reality Visual Displays". In: *IEICE Trans. on Inf. and Syst.* E77-D.12, pp. 1321–1329.

Milleville-Pennel, Isabelle and Camilo Charron (2015). "Driving for Real or on a Fixed-Base Simulator: Is It so Different? An Explorative Study". In: *Presence: Teleoperators and Virtual Environ.* 24.1. doi: 10.1162/PRES_a_00216, pp. 74–91.

Monterrat, Baptiste et al. (Jun. 21-25, 2015). "A Player Model for Adaptive Gamification in Learning Environments". In: *17th Int. Conf. Artificial Intelligence in Education (AIED 2015)*. Vol. 9112. Chapter 30. doi: 10.1007/978-3-319-19773-9_30. Madrid, Spain: Springer, pp. 297–306. ISBN: 978-3-319-19772-2.

Moro, Christian et al. (2017). "The effectiveness of virtual and augmented reality in health sciences and medical anatomy". In: *Anatomical Sci. Educ.* 10.6. doi: 10.1002/ase.1696, pp. 549–559.

Mostafa, Ahmed E et al. (2017). "Designing NeuroSimVR: A Stereoscopic Virtual Reality Spine Surgery Simulator". In: doi: 10.5072/PRISM/31003.

Motejlek, Jiri and Esat Alpay (2018). "A Taxonomy for Virtual and Augmented Reality in Education". In: *Annu. Conf. European Society for Engineering Education*.

Nicolas, Théophane et al. (2015). "Touching and Interacting with Inaccessible Cultural Heritage". In: *Presence: Teleoperators and Virtual Environ.* 24.3. doi: 10.1162/PRES_a_00233, pp. 265–277.

Ochs, M, N Sabouret, and V Corruble (2010). "Simulation of the Dynamics of Nonplayer Characters' Emotions and Social Relations in Games". In: *IEEE Trans. on Computational Intelligence and AI in Games* 1.4. doi: 10.1109/TCIAIG.2009.2036247, pp. 281–297.

Pallavicini, Federica et al. (2013). "Is virtual reality always an effective stressors for exposure treatments? Some insights from a controlled trial". In: *BMC Psychiatry* 13.52. doi: 10.1186/1471-244X-13-52.

Penczek, John et al. (2015). "Evaluating the Optical Characteristics of Stereoscopic Immersive Display Systems". In: *Presence: Teleoperators and Virtual Environ.* 24.4. doi: 10.1162/PRES_a_00235, pp. 279–297.

Pensieri, Claudio and Maddalena Pennacchini (2014). "Overview: Virtual Reality in Medicine". In: *J. For Virtual Worlds Res.* 7.1. doi: 10.4101/jvwr.v7i1.6364.

Poupyrev, I, N Tomokazu, and S Weghorst (1998). "Virtual Notepad: handwriting in immersive VR". In: *IEEE Proc. 1998 Virtual Reality Ann. Int. Symposium.* doi: 10.1109/VRAIS.1998.658467. IEEE Comput. Soc, pp. 126–132. ISBN: 0-8186-8362-7.

Prabhakar, Gowdham et al. (2019). "Interactive gaze and finger controlled HUD for cars". In: *J. on Multimodal User Interfaces* 14.1. doi: 10.1007/s12193-019-00316-9, pp. 101–121.

Ren, Shuo et al. (2015). "Design and Comparison of Immersive Interactive Learning and Instructional Techniques for 3D Virtual Laboratories". In: *Presence: Teleoperators and Virtual Environ.* 24.2. doi: 10.1162/PRES_a_00221, pp. 93–112.

Rickel, Jeff and W Lewis Johnson (May 1, 1998). "STEVE: A Pedagogical Agent for Virtual Reality". In: *Proc. of the Second Int. Conf. on Autonomous Agents (AGENTS '98)*. doi: 10.1145/280765.280851. Minneapolis, MN, USA: ACM Press, pp. 332–333. ISBN: 0897919831.

Roy, Elby, Mahmoud M Bakr, and Roy George (2017). "The need for virtual reality simulators in dental education: A review". In: *The Saudi Dental J.* 29.2. doi: 10.1016/j.sdentj.2017.02.001, pp. 41–47.

Santos, Marc Ericson C et al. (2014). "Augmented Reality Learning Experiences: Survey of Prototype Design and Evaluation". In: *IEEE Trans. on Learning Technologies* 7.1. doi: 10.1109/TLT.2013.37, pp. 38–56.

Schlacht, Irene, Antonio Mastro, and Salman Nazir (Jul. 27-31, 2016). "Advances in Applied Digital Human Modeling and Simulation". In: *Int. Conf. Digital Human Modeling and Simulation (AHFE 2016)*. Walt Disney World, Florida, USA, pp. 1–323.

Schmalstieg, Dieter and Tobias Hollerer (2016). *Augmented Reality: Principles and Practice*. Addison-Wesley. ISBN: 978-0321883575.

Schneider, Matthias et al. (Sep. 22–25, 2019). "A field study to collect expert knowledge for the development of AR HUD navigation concepts". In: *Proc. 11th Int. Conf. on Automotive User Interfaces and Interactive Vehicular Applications Adjunct (AutomotiveUI '19)*. doi: 10.1145/3349263.3351339. Utrecht, Netherlands: ACM Press, pp. 358–362. ISBN: 9781450369206.

Schultze, Ulrike (2010). "Embodiment and presence in virtual worlds: a review". In: *J. of Inf. Technol.* 25.4. doi: 10.1057/jit.2010.25, pp. 434–449.

Shen, Chien-Wen et al. (Apr. 3–7, 2017). "Behavioral Intention of Using Virtual Reality in Learning". In: *Proc. 26th Int. World Wide Web Conf. (WWW'17)*. doi: 10.1145/3041021.3054152. Perth, Australia, pp. 129–137. ISBN: 9781450349147.

Shih, YaChun and MauTsuen Yang (2008). "A collaborative virtual environment for situated language learning using VEC3D". In: *Int. Forum of Educational Technology Society* 11.1, pp. 56–68.

Smith, Matthew, Nigel Stephen Walford, and Carlos Jimenez-Bescos (2019). "Using 3D modelling and game engine technologies for interactive exploration of cultural heritage: An evaluation of four game engines in relation to Roman archaeological heritage". In: *Digit. Appl. in Archaeology and Cultural Heritage* 14. Art no. e00113, doi: 10.1016/j.daach.2019.e00113.

Steuer, Jonathan (1992). "Defining Virtual Reality: Dimensions Determining Telepresence". In: *J. of Communication* 42.4. doi: 10.1111/j.1460-2466.1992.tb00812.x, pp. 73–93.

Sutherland, Ivan E (1965). "The Ultimate Display". In: *Proc. IFIP Congress*.

— (Dec. 9–11, 1968). "A head-mounted three dimensional display". In: *Proc. Fall Joint Computing Conf. (AFIPS '68)*. doi: 10.1145/1476589.1476686. San Francisco, California, USA: ACM Press, pp. 757–764.

Tepper, Oren M et al. (2017). "Mixed Reality with HoloLens". In: *Plastic and Reconstructive Surgery* 140.5. doi: 10.1097/PRS.0000000000003802, pp. 1066–1070.

VanLehn, Kurt (2011). "The Relative Effectiveness of Human Tutoring, Intelligent Tutoring Systems, and Other Tutoring Systems". In: *Educational Psychologist* 46.4. doi: 10.1080/00461520.2011.611369, pp. 197–221.

Våpenstad, Cecilie et al. (2017). "Lack of transfer of skills after virtual reality simulator training with haptic feedback". In: *Minimally Invasive Therapy Allied Technol.* 26.6. doi: 10.1080/13645706.2017.1319866, pp. 346–354.

Vargas, Juan A et al. (May, 2014). "A systematic mapping study on serious game quality". In: *18th Int. Conf. Evaluation and Assessment in Software Engineering (EASE '14)*. 15. doi: 10.1145/2601248.2601261. London, U.K.: ACM Press. ISBN: 9781450324762.

Velev, Dimiter and Plamena Zlateva (2017). "Virtual Reality Challenges in Education and Training". In: *Int. J. of Learn. and Teaching* 3.1. doi: 10.18178/ijlt.3.1.33-37, pp. 1–6.

Vergara, Diego, Manuel Rubio, and Miguel Lorenzo (2017). "On the Design of Virtual Reality Learning Environments in Engineering". In: *Multimodal Technol. and Interaction* 1.2. doi: 10.3390/mti1020011, pp. 11–12.

Vogel, Jennifer J et al. (2006). "Using Virtual Reality with and without Gaming Attributes for Academic Achievement". In: *J. of Res. on Technol. in Educ.* 39.1. doi: 10.1080/15391523.2006.10782475, pp. 105–118.

Warburton, Steven (2009). "Second Life in higher education: Assessing the potential for and the barriers to deploying virtual worlds in learning and teaching". In: *Brit. J. of Educ. Tech.* 40.3. doi: 10.1111/j.1467-8535.2009.00952.x, pp. 414–426.

Ware, Colin, Kevin Arthur, and Kellogg S Booth (1993). "Fish tank virtual reality". In: *Proc. INTERACT Conf. Human Factors in Computing Systems (CHI '93)*. doi: 10.1145/169059.169066. Amsterdam, The Netherlands: ACM Press, pp. 37–42. ISBN: 0897915755.

Westwood, Peter (2008). *What Teachers Need to Know About Teaching Methods*. ACER Press. ISBN: 978-0864319128.

Wilson, Andrew S et al. (2017). "A 3D virtual reality ophthalmoscopy trainer". In: *Clin. Teacher* 14.6. doi: 10.1111/tct.12646, pp. 427–431.

Wilson, M S et al. (1997). "MIST VR: a virtual reality trainer for laparoscopic surgery assesses performance". In: *Ann. of The Royal College of Surgeons of England* 79.6, pp. 403–404.

Witmer, Bob G and Michael J Singer (1998). "Measuring Presence in Virtual Environments - A Presence Questionnaire". In: *Presence: Teleoperators and Virtual Environ.* 7.3. doi: 10.1162/105474698565686, pp. 225–240.

Wolffsohn, J S et al. (1998). "The influence of cognition and age on accommodation, detection rate and response times when using a car head-up display (HUD)". English. In: *Ophthalmic and Physiological Optics* 18.3. doi: 10.1046/j.1475-1313.1998.00350.x, pp. 243–253.

Wynn, Greg, Panagis Lykoudis, and Pasquale Berlingieri (2017). "Development and implementation of a virtual reality laparoscopic colorectal training curriculum". In: *Amer. J. of Surgery* 216.3. doi: 10.1016/j.amjsurg.2017.11.034, pp. 610–617.

Yen, Jung-Chuan, Chih-Hsiao Tsai, and Min Wu (Nov. 26, 2013). "Augmented Reality in the Higher Education: Students' Science Concept Learning and Academic Achievement in Astronomy". In: *13th Int. Educational Technology Conf.* doi: 10.1016/j.sbspro.2013.10.322. Elsevier, pp. 165–173.

You, Fang et al. (Sep. 24–27, 2017). "Using Eye-Tracking to Help Design HUD-Based Safety Indicators for Lane Changes". In: *Proc. 9th Int. Conf. on Automotive User Interfaces and Interactive Vehicular Applications Adjunct (AutomotiveUI '17)*. doi: 10.1145/3131726.3131757. Oldenburg, Germany: ACM Press, pp. 217–221. ISBN: 9781450351515.

Youngblut, Christine (1998). *Educational uses of virtual reality technology*. Tech. rep. IDA-D-2128. Alexandria, VA, USA: Inst. Defense Anal.

Zheng, Min and Mark P Waller (2017). "ChemPreview: an augmented reality-based molecular interface". In: *J. of Molecular Graph. and Model.* 73. doi: 10.1016/j.jmgm.2017.01.019, pp. 18–23.

Zollmann, Stefanie et al. (2020). "Visualization Techniques in Augmented Reality: A Taxonomy, Methods and Patterns". In: *IEEE Transactions on Visualization and Computer Graphics*. doi: 10.1109/TVCG.2020.2986247.